\documentclass[bibyear]{aa}  

\usepackage{natbib,twoopt}
\usepackage[breaklinks=true]{hyperref} 
\bibpunct{(}{)}{;}{a}{}{,}             
\usepackage{graphicx}
\usepackage{txfonts}
\usepackage{hyperref}
\usepackage{booktabs}
\hypersetup{
    colorlinks=true,
    linkcolor=blue,
    citecolor = blue,
    filecolor=magenta,      
    urlcolor=blue}

\usepackage{xcolor}
\usepackage{natbib}
\bibpunct{(}{)}{;}{a}{}{,}
\usepackage{booktabs}
\usepackage{threeparttable}

\def\teff{\textit{$T_{\rm eff}$}}
\def\ali{\textit{A(\rm Li)}}
\def\grav{log\textit{g}}

\begin{document}

\title{The chemical DNA of the Magellanic Clouds }
\subtitle{VII. The lithium abundance in the Small Magellanic Cloud \\
and the discovery of 4 lithium-rich stars.
\thanks{Based on observations collected at the ESO-VLT under the programs 086.D-0665 and 113.26EF.}}

\author{
A. Mucciarelli\inst{1,2} \and
M. Palla\inst{3} \and 
C. T. Nguyen\inst{4,5} \and 
L. Monaco\inst{6}  \and
E. Pancino\inst{3} \and
D. A. Alvarez Garay\inst{3} \and
S. Anoardo\inst{1,7,8} \and \\
C. Lardo\inst{1,2} \and
F. Matteucci\inst{4,5} \and 
D. Romano\inst{2} 
}

\institute{
Dipartimento  di  Fisica  e  Astronomia  “Augusto  Righi”,  Alma  Mater  Studiorum, Universit\`a  di Bologna, Via Gobetti 93/2, I-40129 Bologna, Italy
\and
INAF - Osservatorio di Astrofisica e Scienza dello Spazio di Bologna, Via Gobetti 93/3, I-40129 Bologna, Italy
\and 
INAF – Osservatorio Astrofisico di Arcetri, Largo E. Fermi 5, 50125 Firenze, Italy
\and
University of Trieste, Piazzale Europa, 1, Trieste, Italy
\and 
INAF Osservatorio Astronomico di Trieste, Via Giambattista Tiepolo, 11, Trieste, Italy
\and
Universidad Andres Bello, Facultad de Ciencias Exactas, Departamento de F{\'i}sica y Astronom{\'i}a - Instituto de Astrof{\'i}sica, Autopista
Concepci\'on-Talcahuano 7100, Talcahuano, Chile
\and
Leibniz-Institut f\"ur Astrophysik Potsdam, An der Sternwarte 16, D-14482 Potsdam, Germany 
\and 
Institut f\"ur Physik und Astronomie, Universit\"at Potsdam, Haus 28, Karl-Liebknecht-Str. 24/25, D-14476 Potsdam, Germany
}

\date{Received XX XX, 2026}

\abstract{Measuring lithium abundance, \ali , in galaxies beyond the Milky Way remains a challenging task. 
In these systems, for stellar populations older than $\sim$1 Gyr, only stars close to the tip of the red giant branch (RGB) are sufficiently bright 
to be observed spectroscopically. Their \ali\ are expected to be significantly altered 
by mixing episodes that occur during the stellar evolution, making the interpretation of measured abundances more complex.

In this context, we present \ali\ of 209 RGB stars in the Small Magellanic Cloud (SMC). 
The observations were performed with FLAMES at the VLT.  Approximately one-third of the sample
shows a lithium line detected at a significance level greater than 5$\sigma$, while about half of the sample has detections 
above 3$\sigma$; for the remaining stars, only upper limits can be derived. 
As a result, this sample exhibits a significant star-to-star scatter in A(Li).

Using theoretical models, we find that the observed scatter can be likely explained by the combination 
of two effects. 
First, the stars in our sample span a wide range of ages, and therefore stellar masses, as well 
as metallicities. Thermohaline mixing, one of the leading mechanisms proposed to explain the extra mixing occurring 
at the RGB bump, predicts substantially different levels of lithium depletion depending on stellar mass and metallicity, 
naturally contributing to the observed scatter in A(Li).
Second, in the SMC, the enrichment in A(Li) driven by novae production is expected to occur at lower metallicities than 
in our Galaxy, because the SMC has experienced a lower star formation efficiency than the Milky Way. 
This is also supported by the increase of \ali\ with decreasing stellar age, as measured in three SMC globular clusters. 

Furthermore, we report the first identification of four low-mass Li-rich giant stars in the SMC, reinforcing the existence 
of Li-rich stars in extra-galactic systems. All these results demonstrate 
that lithium abundances measured in stars close to the RGB tip,
still provide valuable constraints on both stellar and galactic chemical evolution of lithium in external galaxies.
}


\maketitle
%

\section{Introduction}

Lithium (Li) remains one of the most intriguing chemical elements to study, as its production is broadly linked 
to primordial Big Bang nucleosynthesis, stellar evolution, and contributions from cosmic rays and novae. 
Its main isotope, $\rm ^{7}Li$, is produced directly in the Big Bang nucleosynthesis in the first 
minutes of the Universe \citep[see e.g.][]{peebles1993,fields2020}. 
In the subsequent stellar generations,  $\rm ^{7}Li$ produced 
from decay of $\rm ^{7}Be$ is immediately destroyed at relatively low 
temperature ($\sim2.5\cdot10^{6}$ K) through $\rm ^{7}Li(p,\alpha) ^{4}He$ reaction. 
This fragility leads to a significant drop of the surface Li abundance A(Li)\footnote{A(Li) is defined as $\log{\frac{\rm N_{Li}}{\rm N_{H}}}+12$} 
whenever the convective envelope deepens into the stellar interior, 
reaching stellar layers that are hot enough to burn Li. For instance, 
in low-mass stars A(Li) drops 
at the first dredge-up \citep[see e.g.][]{iben1967} and at the Red Giant branch (RGB) bump \citep{gratton2000,lind2009}.

When we consider a specific evolutionary sequence, A(Li) in our Galaxy follows a characteristic and well-defined trend. 
For instance, among the unevolved stars with [Fe/H] between --2.8\,dex and --1.0\,dex, A(Li) remains constant to a value of A(Li)$\sim$2.1-2.3 dex 
\citep[the so-called Spite plateau,][]{spite1982,bonifacio1997,charbonnel2005,bonifacio2007}, 
while the RGB pre-bump stars have a constant value around 0.9-1.0 dex \citep{mucciarelli2012,mucciarelli2022}.

For decades the standard interpretation of the Spite plateau was that the measured abundance reflects the 
abundance of Li produced in the Big Bang nucleosynthesis. This view is in conflict with the value of A(Li) 
derived from standard Big Bang nucleosynthesis and the baryon-to-photon ratio determined by the
Wilkinson Microwave Anisotropy Probe and Planck satellites \citep[A(Li)$\sim$2.7 dex,][]{Pitrou18}. The two values can be reconciled each other invoking 
additional element transport mechanisms occurring during the main sequence evolution \citep[see e.g.][]{richard2005,korn2006,deal2021,mucciarelli2022}.

At higher metallicities than those of the Spite plateau, A(Li) starts to increase up 
to the meteoritic value  \citep[\ali\ =3.33 dex,][]{Lodders09} at solar metallicity and exhibiting a large star-to-star scatter. 
In fact, additional sources of Li production may operate on long timescales (several Gyr), thus providing fresh 
Li when [Fe/H] is already high ([Fe/H]$>$--0.8 dex). 
These sources include novae \citep{starrfield,dantona91}, asymptotic giant branch (AGB) stars \citep{cameron71}, 
$\nu$ process during core-collapse supernovae explosions \citep{matteucci1995} and high-energy Galactic cosmic rays \citep{reeves70,prantzos2012}. 
Novae in particular are considered an efficient source of Li and are able to explain the rise in A(Li) observed 
in thin disk stars \citep{romano99,grisoni2019,romano21}. 
Usually the upper envelope of  A(Li) distribution in the thin disk star is considered as diagnostic of the 
true A(Li) \citep[see e.g.][]{Cescutti19,romano21,nguyen25}, while the large A(Li) star-to-star scatter 
observed at any evolutionary stage is usually interpreted as depletion caused by stellar interior mixing processes.

A further complication of this complex picture is the existence of a small number of stars with A(Li) 
that are significantly higher than those of most of the stars at the same evolutionary stage and metallicity. 
The origin of these rare Li-rich stars \citep[$\sim$1\%,][]{deepak2019,martell2021} is still debated and they could be explained 
with internal or external processes. Internal processes require the onset of the Cameron-Fowler mechanism 
\citep{cameron55,cameron71} able of preserving the $^{7}{\rm Li}$ produced by the decay of $\rm ^{7}Be$. 
This mechanism can work in low-mass stars whenever a mechanism carries $\rm ^{7}Be$
to cooler stellar regions. 
An external channel to create Li-rich stars is the mass transfer process in a binary system 
where the companion star is able to produce fresh Li through the Cameron-Fowler mechanism. 
This scenario is usually invoked to explain the existence of metal-poor, low-mass Li-rich stars before the RGB bump 
where no mechanism are able to onset the Cameron-Fowler mechanism. 
Another possible mechanism, able to explain the most metal-rich Li-rich stars, is the engulfment of small bodies, 
i.e. rocky planets, hot Jupiters, and brown dwarfs \citep{siess99}.

A(Li) in external galaxies, in turn, is crucial for understanding (i)~whether these systems were born 
with the same initial A(Li) measured in the Milky Way (MW), and (ii)~the role of novae and sources of Li production 
to the chemical evolution history of galaxies with star formation efficiencies differing from that of the MW. 
At present, our knowledge of A(Li) outside our Galaxy is extremely limited 
because of the challenging of these observations. 
For stellar populations older than $\sim$1-2 Gyr, the only observable stars are those near the RGB tip, 
which have already experienced extra-mixing episodes.
Such mixing dramatically reduces the surface \ali, making the Li resonance line at 6707.8 \AA\ virtually 
absent or undetectable. On the other hand, for younger stellar populations, 
Li measurements can be carried out, in particular in bright AGB stars and carbon stars 
\citep[see e.g.][]{plez1993,smith1995,Hatzidimitriou2003}, providing information about the recent 
Li enrichment in external galaxies with an ongoing star formation, e.g. the Magellanic Clouds. 
Finally, it is worth to mention that interstellar \ali\ and  $^{6}\mathrm{Li}/^{7}\mathrm{Li}$ 
isotopic ratio in the Small Magellanic Cloud (SMC) 
have been derived from spectra along the line of sight toward Sk 143 \citep{howk12,molaro24}, 
adding another effort to study Li in this galaxy.

In this paper, we present the measure of A(Li) for the first time in a large sample of post-RGB bump stars in the SMC 
observed with high-resolution optical spectra. 
Stellar and chemical evolution models were computed to comprehend the observing trends and findings.
The paper is structured as follows: Section~\ref{obs} describes the spectroscopic dataset, 
Section~\ref{chems} describes the determination of the surface \ali, while 
Section~\ref{lithium} discusses the derived distribution of A(Li) as a function of the 
effective temperature and metallicity. The chemical and stellar evolution models used 
to interpret the results are presented in Section~\ref{models}. 
Four targets with anomalous A(Li) and identified as Li-rich stars are discussed 
in Section~\ref{lirich}. Finally, Section~\ref{end} summarizes the main results and conclusions.

\section{Observations}
\label{obs}
This study is based on observations performed with the multi-object 
spectrograph FLAMES \citep{pasquini} at the Very Large Telescope of the European Southern Observatory. 
The dataset is composed by: \\
(a)~209 giant stars members of the SMC observed with GIRAFFE-MEDUSA HR15N setup 
(ID Program: 113.26EF, PI: Mucciarelli), with a spectral resolution of 19\,200 and ranging from 
6470 \AA\ to 6790 \AA , sampling the resonance Li line at 6707.8 \AA\ . 
This sample is the same already discussed in \citet{anoardo}, with a large overlap (158 targets in common) 
to the sample by \citet{muc23a}. The two studies provided for these 209 stars 
atmospheric parameters, radial velocity (RV) and chemical abundances for the main groups of elements 
(i.e. $\alpha$, iron-peak, slow and rapid neutron capture elements).  
These stars are located in three different SMC regions, two fields (FLD~339 and FLD~419) in the SMC old spheroid, 
and one (FLD~121) in the outskirts of the galaxy. 
For each field, three exposures of 45 minutes each were secured. 
The signal-to-noise (S/N) ratio per pixel around the Li line  for the co-added spectra  ranges from $\sim$50 to $\sim$120; \\
(b)~14 giant stars members of three SMC globular clusters (GCs) observed with the FLAMES-UVES spectrograph 
adopting the Red Arm 580 UVES setup with spectral resolution of R=~47\,000 and
spectral coverage 4760–6840 \AA\ (ID program 086.D-0665, PI: Mucciarelli). 
The three GCs were discussed in \citet{muc23b} that derived their chemical abundances (except for Li) 
and they sample different ages of the life of the SMC: NGC~121 is an old cluster \citep[$\sim$10.5 Gyr,][]{glatt2008a}, 
NGC~339 has $\sim$6 Gyr \citep{glatt2008b} and NGC~419 has  $\sim$1.5 Gyr \citep{glatt2008b}. 
5, 4 and 5 member stars were observed for NGC~121, NGC~339 and NGC~419, respectively.
A total of 12, 14 and 10 exposures of 45 min each are secured for NGC~121, NGC~339 and NGC~419, respectively, 
securing S/N ratios of about 30-40 around the Li line. 

The spectra were reduced with the FLAMES-GIRAFFE and FLAMES-UVES standard pipelines that include
bias-subtraction, flat-fielding, spectral extraction, wavelength calibration, and, only for FLAMES-UVES spectra, 
the order merging. The sky subtraction was performed by subtracting a master sky spectrum obtained as 
the median-averaged spectrum of $\sim$15 sky spectra for FLAMES-GIRAFFE and the mean of 2 sky spectra for FLAMES-UVES.

\section{Chemical analysis}
\label{chems}
\subsection{Line detection}
As a first step, we checked in each spectrum whether the resonance Li line at 6707.8 \AA\ is detectable. 
In about half of the FLAMES-GIRAFFE spectra, no spectral feature is present in the vicinity of the wavelength of the Li line, while in other spectra 
absorption lines with different strengths are clearly visible.
In order to decide, for each spectrum, whether the Li line is detected at a statistically significant level, we adopted the following procedure. 
Two local fits were performed in a spectral window centred on the expected position of the Li line and broad $\pm$3 
Full Width Half Maximum (FWHM). 
The first fit assumes a pure continuum model, corresponding to the null hypothesis of no line being present. 
The second fit includes a Gaussian profile describing a possible absorption line, with the continuum level, the line centre and the FWHM fixed to their expected values, 
while leaving the line amplitude as a free parameter.
We then computed the difference between the $\chi^2$ values obtained for the two models, 
$\Delta\chi^2 = \chi^2_{\mathrm{cont}} - \chi^2_{\mathrm{line}}$ .
Typically, in order to decide whether the model including the line provides a better representation of the data than the continuum-only model, Wilks theorem is adopted \citep{Wilks1938}, 
according to which $\Delta\chi^2$ follows a $\chi^2$ distribution with one degree of freedom (corresponding to the number of additional free parameters in the second model, in this case only the line amplitude), 
$\Delta\chi^2\sim\chi^{2}(1)$ .

In the case of an absorption line, the Wilks theorem cannot be used because 
the line amplitude is constrained to be non-negative (absorption only) and the null hypothesis lies (continuum model) on the boundary of the parameter space \citep[see e.g.][]{Protassov02}.
As a consequence, the likelihood-ratio statistic does not follow Wilks theorem, but instead follows the Chernoff distribution 
\citep{Chernoff1954} corresponding to a 
50–50 mixture of a Dirac delta function at zero and a $\chi^2$ distribution with one degree of freedom, 
$\Delta\chi^2\sim\frac{1}{2}\delta(0)+\frac{1}{2}\chi^{2}(1)$.
The statistical significance that the model including the Li line provides an improvement over the pure continuum model was then evaluated accordingly, 
Z = $\Phi^{-1}(1 - p)$, where the one-sided p-value is 
p = $\tfrac{1}{2}\, P\!\left(\chi^2_{1} \ge \Delta\chi^2\right)$ and 
$\Phi$ is the inverse function of the normal cumulation distribution.

\begin{figure}
  \includegraphics[width=0.5\textwidth]{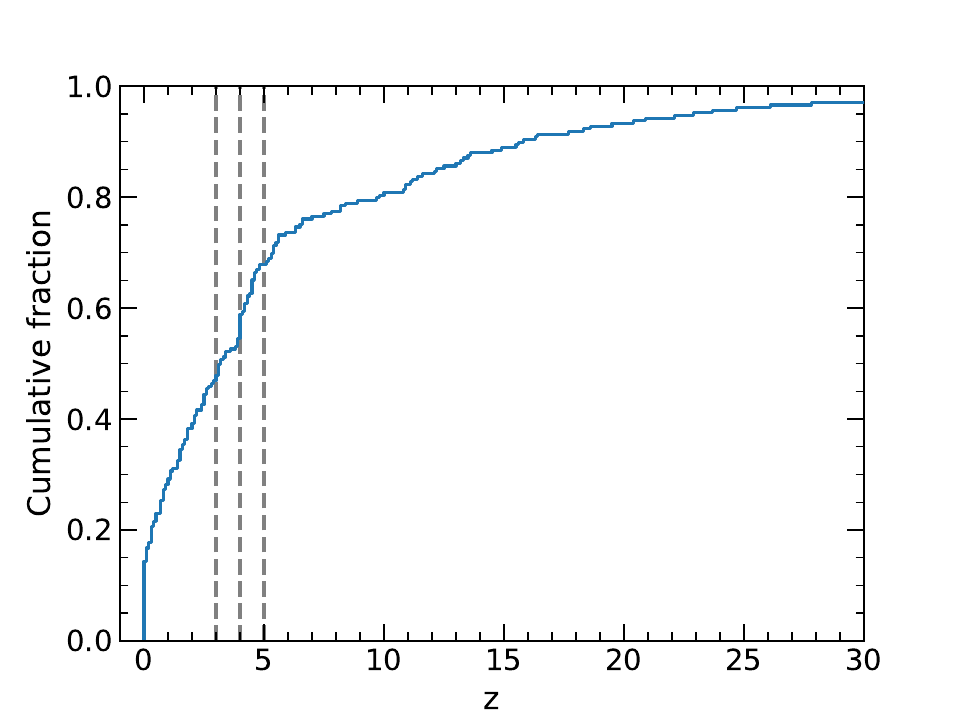}
  \caption{Cumulative distribution of the 
  statistical significance Z to detect the Li line at 6707.8 \AA\ in the spectroscopic FLAMES-GIRAFFE sample. 
  The three vertical dashed lines indicate Z values of 3, 4 and 5.}
  \label{fig_cumdis}
\end{figure}

Fig.~\ref{fig_cumdis} shows the cumulative distribution of the statistical significance Z to detect Li line in the entire 
FLAMES-GIRAFFE sample. The Li line can be detected at a level higher than 3, 4 and 5$\sigma$ in 111, 95 and 67 stars, respectively, 
corresponding to a percentage of 53, 45, and 32\% of the entire sample.  

\subsection{\ali\ determination}

\begin{figure}
  \includegraphics[width=0.53\textwidth]{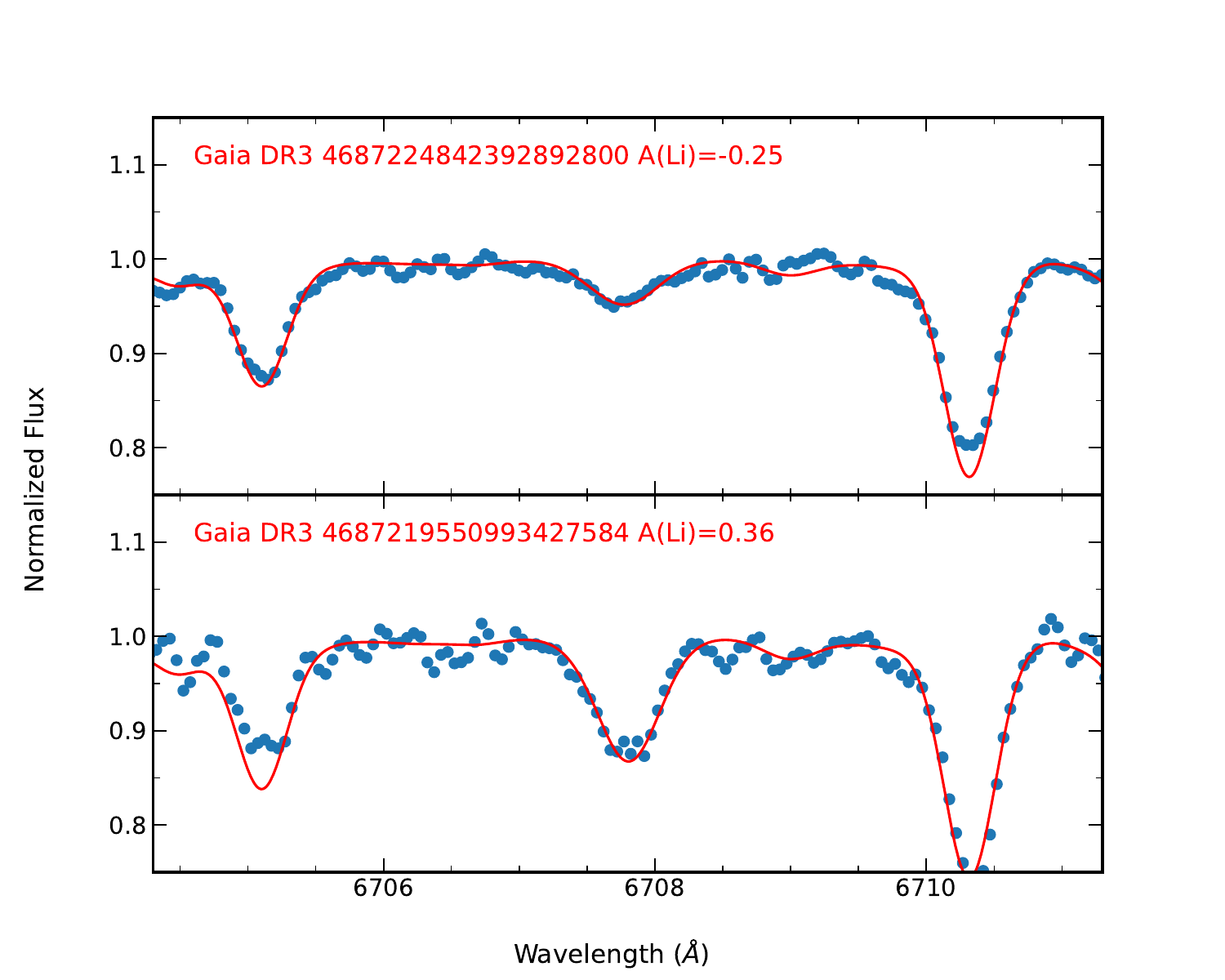}
  \caption{Spectral regions around the Li line at 6707.8 \AA\ for the target stars 
  Gaia DR3 4687224842392892800 (upper panel), and Gaia DR3 4687219550993427584 (lower panel, blue points), 
  where the Li line has been detected with a significance Z$>$5, 
  with superimposed the best-fit synthetic spectra (red lines). }
  \label{fit1}
\end{figure}

Secure detections are defined as features with a Gaussian-equivalent significance 
Z$\ge$5, computed using the Chernoff distribution as described above. 
For these stars the corresponding A(Li) was measured by matching the line profile with synthetic spectra 
through a $\chi^2$-minimization using the code SALVADOR (Alvarez Garay et al. in prep.). 
The synthetic spectra were calculated with the code SYNTHE \citep{kurucz2005} adopting the appropriate 
stellar parameters \citep[see][]{anoardo}, the latter computed using the Gaia Data Release 3 photometry 
and the colour-$\rm T_{eff}$ relations by \citet{muc21}. 
The adopted model atmospheres were calculated with the code ATLAS9 \citep{kurucz70,kurucz2005}
and using the new opacity distribution functions available in the KOALA database \citep{koala}. 
The linelist of  Li line, including hyperfine and isotopic splitting, is from \citet{yan95}. 
In the calculation of the synthetic spectra we adopted the solar isotopic ratio $\rm ^{6}Li/^{7}Li$=0.08 but 
this assumption has no impact on the derived abundance. Only for the four Li-rich stars identified in the sample (see Section~\ref{lirich}), 
we adopted  $\rm ^{6}Li/^{7}Li$=0.0 to improve the quality of the  best-fit because of the difficulty to 
simultaneously reproduce the line depth and broadening of such strong (EW$>$200  m\AA ) Li lines \citep[see e.g.][]{monaco2014, mucciarelli21}.
Abundance uncertainties were computed by summing in quadrature the main sources of uncertainties, 
namely those arising from the atmospheric parameters and that from the fitting procedure. 
The latter were estimated with Monte Carlo simulations and repeating for each star the analysis on a 
sample of 300 synthetic spectra where appropriate Gaussian noise was injected. 
Fig.~\ref{fit1} shows examples of best-fit synthetic spectra obtained for two stars 
with a Gaussian-equivalent significance Z$\ge$5 and two extreme value of A(Li).

For all the stars with Z$<$3 (secure non-detections of the Li line), we estimated an upper limit 
of A(Li). We calculated the largest equivalent width (EW) detectable within the local noise as three times 
the uncertainty associated to the measured EW according to the Cayrel formula \citep{cayrel88}. 
For each star an appropriate curve of growth for the Li line was calculated by integrating the 
synthetic line profiles obtained with different A(Li). 
These curves of growth were used to derive the A(Li) associated to 3$\sigma_{EW}$. 

\begin{figure}
  \includegraphics[width=0.5\textwidth]{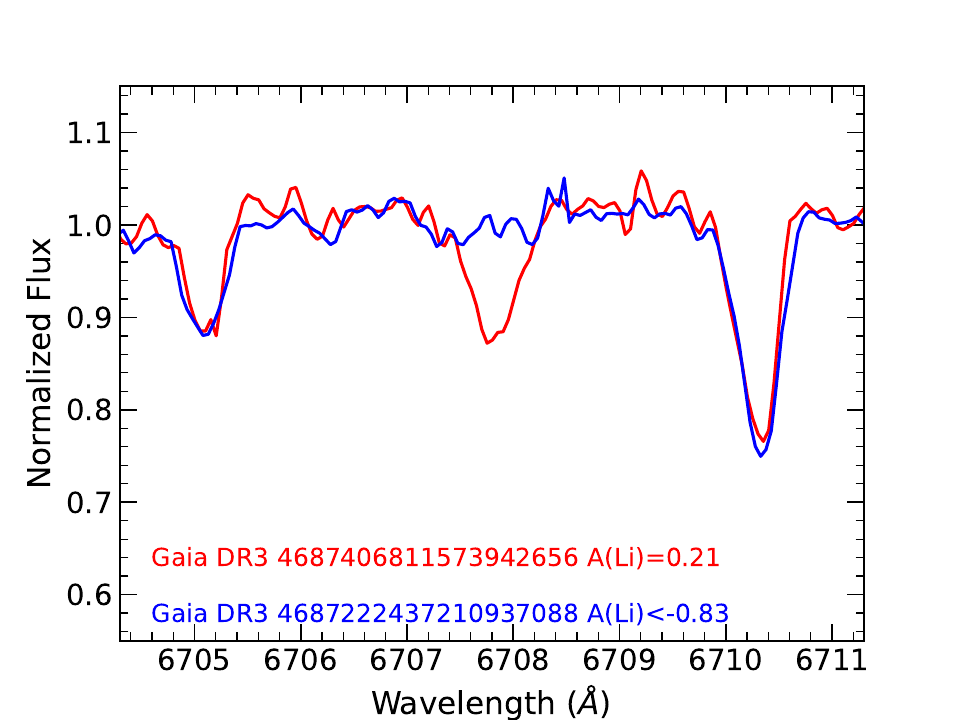}
  \caption{Comparison between the spectral region around Li line for two spectra 
  with similar parameters and metallicities but different statistical significance Z
  of Li line: Gaia DR3 4687406811573942656 with $\rm T_{eff}$=~3996 K, log~g=0.57, 
  [Fe/H]=--1.02 dex, Z=23.7 (red spectrum) and Gaia DR3 4687222437210937088 with
  $\rm T_{eff}$=~4007 K, log~g=0.42, 
  [Fe/H]=--1.04 dex, Z=2.5 (blue spectrum).}
  \label{spectrum1}
\end{figure}

Lower-significance (Z=3–5) features are classified as candidate stars with a detection of the Li line 
but are not considered secure detections. 
For the stars flagged as candidates, we adopted the abundance derived from the corresponding 
curve of growth and the measured EW.
For both true measurements and upper limits, the derived LTE A(Li) were corrected using the 
3D-NLTE corrections by \citet{wang21} (and listed in Table~\ref{info}).
Figure~\ref{spectrum1} shows the comparison between two spectra of stars with similar \teff\ and [Fe/H] 
for which we provide an upper limit and a true measurement.  
Table~\ref{info} lists all the spectroscopic targets, 
the derived \ali , together with the iron abundance [Fe/H] and 
the atmospheric parameters.

\begin{table*}[h!]
\caption{Information about the SMC spectroscopic targets.}
\label{info}
\centering
\begin{threeparttable}
\begin{tabular}{cccccccc}
\toprule
ID Gaia DR3 & [Fe/H] &  \teff\ & \grav\ & \ali & $\sigma_{\textit{A(\rm Li)}}$ & $\delta{\textit{A(\rm Li)}}$ & {\rm flag} \\
\midrule
   & (dex)  &  (K)  &  (dex)  & (dex)  & (dex) & (dex) & \\
\midrule
4689845798923576704  &   $-$1.01   &  4319  &   1.06   &   $-$0.24   &    ---   &  $-$0.04   &   0 \\
4689845597059733248  &   $-$0.89   &  4065  &   0.84   &   $-$0.26   &    0.15   &     0.03   &   1 \\
4689858172724094720  &   $-$1.17   &  4084  &   0.88   &   $-$0.59   &    ---   &     0.02   &   0 \\
4689852189834915072  &   $-$1.14   &  4345  &   1.09   &   $-$0.18   &    ---   &  $-$0.05   &   0 \\
4689844978584591104  &   $-$0.82   &  4293  &   1.12   &   $-$0.20   &    ---   &  $-$0.03   &   0 \\
4689846730930989440  &   $-$0.98   &  4029  &   0.63   &     +0.87   &    0.12   &     0.02   &   2 \\
4689859031717528576  &   $-$1.75   &  4424  &   1.09   &   $-$0.07   &    ---   &  $-$0.10   &   0 \\
4689862914367958144  &   $-$0.99   &  4194  &   0.93   &   $-$0.28   &    ---   &     0.00   &   0 \\
4689851266416720768  &   $-$1.09   &  4093  &   0.63   &   $-$0.69   &    ---   &     0.02   &   0 \\
4689864078304620544  &   $-$0.81   &  4084  &   0.77   &   $-$0.60   &    ---   &     0.02   &   0 \\

\bottomrule
\end{tabular}
\begin{tablenotes}
\small
\item \textbf{Notes.} Columns are: ID from Gaia DR3 \citep{vallenari23}, 
[Fe/H], \teff\ and log~g from \citet{anoardo}, 3D-NLTE \ali\ and corresponding error, 
3D-NLTE correction by \citet{wang21}, flag to indicate 
upper limit (0), candidate (1) and true measurements (2).
The entire table is available at the CDS.
\end{tablenotes}
\end{threeparttable}
\end{table*}

\section{A(Li) distribution} 
\label{lithium}

\subsection{SMC field stars}

We compared \ali\ in the SMC field stars with those 
in the MW stars measured in the GALAH Data Release 4 \citep[DR4,][]{buder25}. 
We adopted the value-added catalogue galah\_dr4\_vac\_3dnlte\_a\_li\_240705.fits 
that includes \ali\ derived following \citet{wang21,wang2024}, therefore 
on the same 3D-NLTE scale of this study.
In more details, from the GALAH DR4 catalogue, we selected stars  with \teff\ $<$5000 K and 
log~g$<$2.0 to be consistent with the evolutionary stage of the targets of our sample.
Approximately 60\% of the post-RGB bump stars in GALAH DR4 sample have only upper limits on \ali\ . 
For the remaining stars, \ali\ is reported as a true measurement. 

A detailed check of the GALAH DR4 results and spectra revealed that a large amount of stars classified as true measurements have extremely small EWs  
the Li line ($\sim$60\% with EW$<$10 m\AA\ and $\sim$70\% with EW$<$20 m\AA ), 
whose reliable detection appears questionable given the S/N ratio
distribution of the GALAH spectra and the metallicity of the stars, most of which are of solar or super-solar metallicity. 
In such metal-rich stars, line blending and continuum placement uncertainties are expected to significantly affect the detectability of very weak Li features. 
Appendix \ref{app:galah} provides more details about the \ali\ distribution in post-RGB bump stars as measured in GALAH DR4.
For the comparison presented below, we therefore adopted a conservative approach and treated stars flagged 
as true measurements but with EW$<5\sigma_{Cayrel}$ as upper limits (see Section~\ref{lirich} for details). This choice 
ensure robust Li line detections in our sample and minimize the risk of false positives.
reducing the risk of misinterpreting the A(Li) distribution of 
MW stars and provides a more robust reference sample for comparison with our targets.

\begin{figure}
  \includegraphics[width=0.52\textwidth]{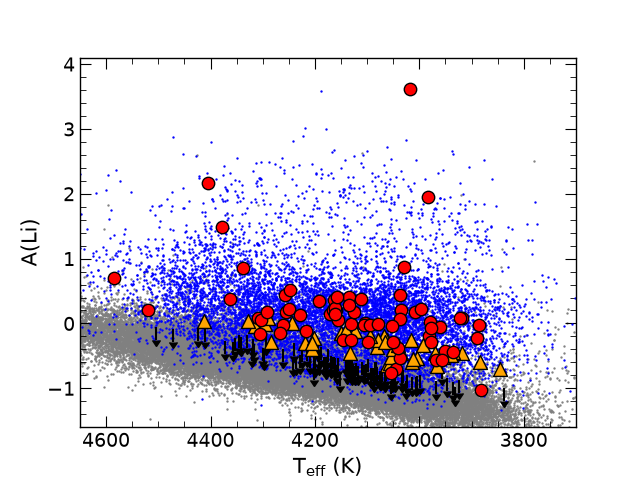}
  \includegraphics[width=0.52\textwidth]{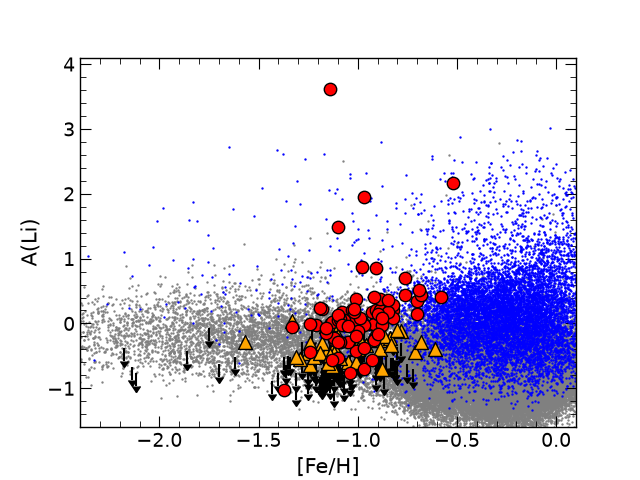}
  \caption{Behaviour of \ali\ as a function of \teff\ (upper panel) and [Fe/H] (lower panel) for the SMC field 
   stars analysed in this study: stars with a safe detection of the Li line are plotted as red circles, candidate stars 
   with a detection of Li line with Z=3-5 are plotted as orange triangles, and 
  the upper limits are plotted as black arrows. Small points are the post-RGB bump MW stars from GALAH DR4: 
  blue points are true measurements with EW of Li line larger than 5$\sigma_{Cayrel}$ , while 
  grey points are measures with EW smaller than $5\sigma_{Cayrel}$ and upper limits.}
  \label{fig_lidstr}
\end{figure}

The entire spectroscopic sample of SMC field stars covers a large range of \ali\ values, 
with the true measurements ranging from $\sim$--1 up to $\sim$+3.6 dex. 
Fig.~\ref{fig_lidstr} shows the behaviour of \ali\ as a function of [Fe/H] and \teff\ , 
highlighting true measurements, candidates and upper limits.
The \ali\ distribution of the SMC field stars exhibits a large star-to-star scatter that cannot 
be explained solely by the uncertainties in \ali\ , 
which are typically of the order of 0.1–0.15 dex. 
In the \teff\--\ali\ plane (upper panel of Fig.~\ref{fig_lidstr}), 
the SMC stars for which true A(Li) measurements are available overlap with the bulk of the MW population when considering only stars 
with genuine Li detections and EW$>5\sigma_{Cayrel}$.
On the other hand, the SMC stars for which only upper limits on A(Li) are available occupy the same average locus as MW stars 
with upper limits or with EW $<5\sigma_{Cayrel}$. Four stars stand out as having \ali\ $>$1 dex, significantly higher than the bulk of the sample 
(see Section~\ref{lirich} for more discussions).

The comparison between the SMC and MW samples in the [Fe/H]–-\ali\ plane (bottom panel of Fig.~\ref{fig_lidstr}) reveals an important difference. 
The MW stars with measured \ali\ have almost exclusively 
[Fe/H]$>$--0.5 dex. This implies that this is substantially more metal-rich than the SMC population. 
This suggests that, despite the similarity between the two samples in \teff\--\ali\ plane, 
they show a remarkable difference in metallicity distribution.

\begin{figure}
  \includegraphics[width=0.52\textwidth]{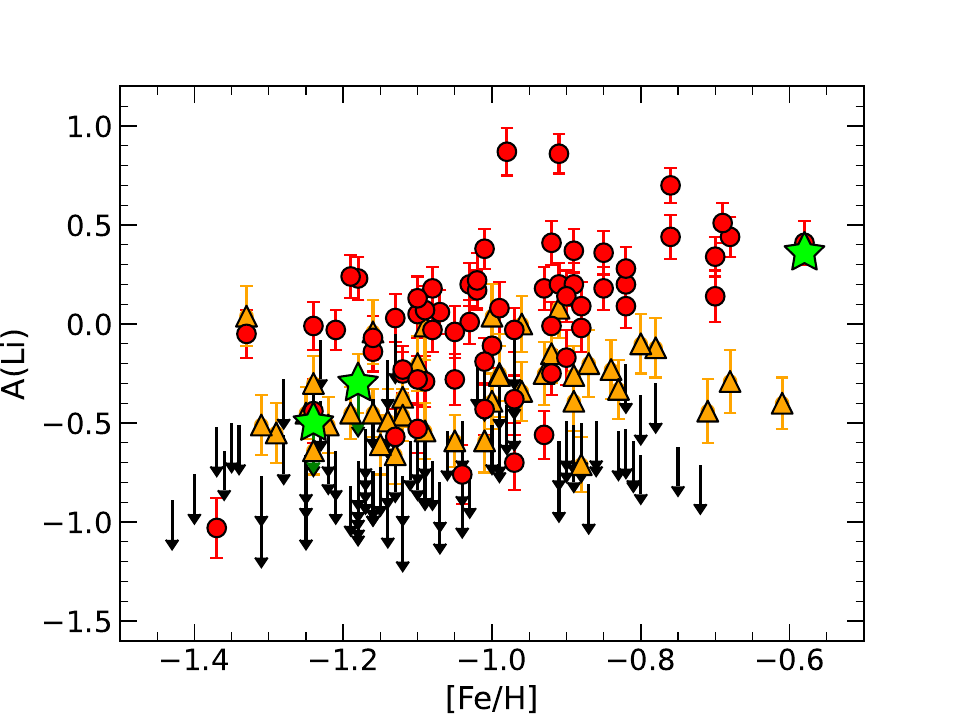}
  \caption{Behaviour of \ali\ as a function of [Fe/H] 
  in the SMC field for stars within [Fe/H]$\sim$--1.4 to --0.6 dex and A(Li)$<$+1 dex 
  (same symbols as Fig.~\ref{fig_lidstr}).
  Green star-symbols indicate the results for the three SMC GCs.}
  \label{fig_lidstr2}
\end{figure}

Finally, the FLAMES-GIRAFFE sample seems to suggest an increase of \ali\ with [Fe/H] (see Fig.~\ref{fig_lidstr2}).
A Pearson correlation test provides a p-value of $\sim$0.003 for both sub-samples (with Z$>$3 and with Z$>$5), 
pointing out that the observed increase of \ali\ with [Fe/H] is real.

\subsection{SMC globular clusters} 

Although the available sample includes only three SMC GCs, it provides a first indication 
of the temporal evolution of A(Li) in this galaxy, since the ages of these clusters are well constrained \citep[see e.g.][]{glatt2008a,glatt2008b, dalex2016}, 
whereas those of field stars are much more uncertain.
In the FLAMES-UVES spectra of the three SMC GCs, the Li line is clearly detectable 
(with statistical significance Z larger than 10) only in the spectra of 4 (out 5) stars of the young, metal-rich cluster NGC~419. 
In contrast, in the other two clusters, the Li line is not detectable in any of the spectra and we provide only 
upper limits for A(Li). Table~\ref{infogc} lists the derived \ali\ for each individual cluster stars.
For each cluster, we adopt as the final A(Li) value the mean of the four individual measurements for NGC~419, 
where Li is detected, and the lowest upper limit for the other two clusters, corresponding to the star with the highest S/N ratio.

\begin{figure}
  \includegraphics[width=0.52\textwidth]{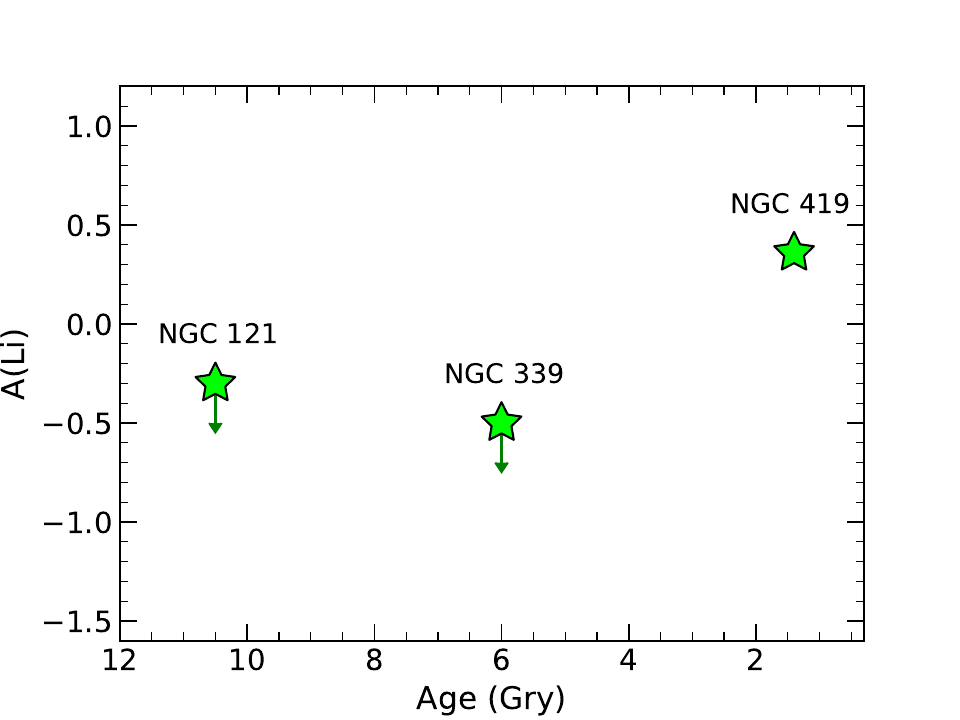}
  \caption{Behaviour of \ali\ as a function of the ages for the 
  three SMC globular clusters of the sample.}
  \label{amr}
\end{figure}

Fig.~\ref{fig_lidstr2} shows the behaviour of \ali\ in these three GCs as a function of [Fe/H]: 
NGC~121 and NGC~339, both with [Fe/H]$\sim$--1.2 dex and ages older than 6 Gyr have \ali\ about 
lower than --0.3 and --0.5 dex, respectively , while the metal-rich and young cluster NGC~419 exhibits a higher \ali\, with an average value 
+0.36$\pm$0.04 dex ($\sigma$=0.09 dex). This increase of \ali\ with [Fe/H] corresponds also to an increase of \ali\ 
by decreasing the age, as shown in Fig.~\ref{amr}.

\begin{table*}[h!]
\caption{\ali\ for individual stars in three SMC GCs.}
\label{infogc}
\centering
\begin{threeparttable}
\begin{tabular}{cccc}
\toprule
Cluster & ID & \ali & $\sigma_{\rm A(Li)}$  \\
\midrule
   &  &  (dex)  &  (dex)  \\
\midrule
NGC~121 & 9 & $<-$0.3  & -- \\
NGC~121 & 14 & $<-$0.3  & -- \\
NGC~121 & 18 & $<-$0.2  & -- \\
NGC~121 & 31 & $<-$0.2  & -- \\
NGC~121 & 35 & $<-$0.1  & -- \\
\hline
NGC~339 & 535 & $<-$0.4  & -- \\
NGC~339 & 893 & $<-$0.5  & -- \\
NGC~339 & 958 & $<-$0.2  & -- \\
NGC~339 & 1076 & $<-$0.2  & -- \\
\hline
NGC~419 & 727  & 0.33 & 0.15 \\
NGC~419 & 732  & 0.48 & 0.12 \\
NGC~419 & 852  & 0.37 & 0.13 \\
NGC~419 & 1384 & $<$0.30  & -- \\
NGC~419 & 1633 & 0.26 & 0.13 \\
\hline
\bottomrule
\end{tabular}
\begin{tablenotes}
\small
\item \textbf{Notes.} Columns are: name of the SMC GCs, 
ID of the star target from \citet{muc23b},  \ali\ and corresponding error. 
Additional information about the stellar parameters and [Fe/H] of each star 
are in \citet{muc23b}.
\end{tablenotes}
\end{threeparttable}
\end{table*}

\section{The scatter of the A(Li) in the SMC: possible scenarios}
\label{models}

The main feature in the \ali\ distribution of the SMC is the observed abundance spread, not 
ascribable to the abundance uncertainties, as clearly visible in Fig.~\ref{fig_lidstr2}.
The observed scatter is likely the result of the combined effect of two main physical processes operating in these stars: 
the mass-dependent efficiency of thermohaline mixing, being a likely mixing mechanism after the RGB bump, and 
the Li enrichment from novae. In the following we try to evaluate these two mechanisms 
using appropriated theoretical models. Additionally, we discuss the role played by another effect, the 
stellar rotational velocity, that could increase the \ali\ scatter.

\subsection{Lithium depletion during the post-RGB bump phase} 

\begin{figure*}
  \includegraphics[width=2\columnwidth]{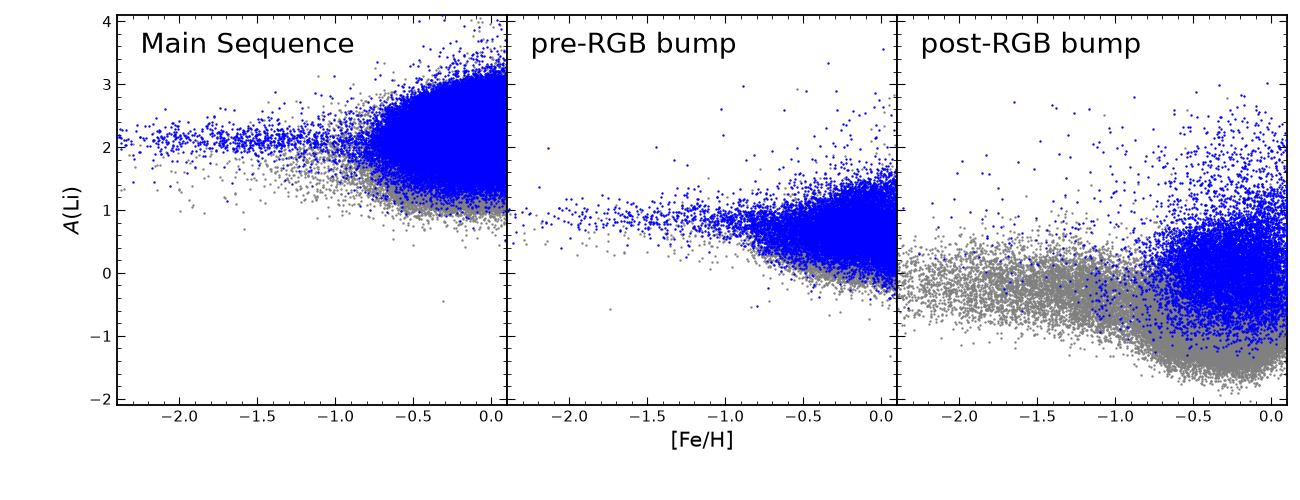}
  \caption{Behaviour of \ali\ as a function of [Fe/H] for MW stars of GALAH DR4 \citep{wang2024} selected according to 
  their evolutionary stage: main sequence stars (left panel, \teff\ =5850-6800 K, log~g=3.6-4.5),
  pre-RGB bump stars (middle panel,\teff\ =4600-5200 K, log~g=3.0-3.5), 
  and post-RGB bump stars (right panel, \teff\ $<$5000 K, log~g$<$2.0).
  Blue and grey circles indicate true measurements and upper limits, respectively.}
  \label{galah_fig}
\end{figure*}

The mixing that occurs after the RGB bump not only significantly reduces the \ali, but also increases the star-to-star scatter 
in \ali. Figure~\ref{galah_fig} shows \ali\ as a function of [Fe/H] for GALAH DR4 stars 
selected in the main sequence and in the RGB (before and after the bump).
For stars with [Fe/H]\,$<$\,-0.8 dex, where the contribution of novae to Li production 
is not yet significant, the three samples exhibit different A(Li) values, with a larger star-to-star scatter in the post-RGB bump sample. 
As will be discussed below, the efficiency of the mixing mechanism responsible for reducing A(Li) is expected to depend on both stellar mass and metallicity, 
therefore introducing a relevant star-to-star scatter in samples of stars with different ages and [Fe/H], like in the SMC 
discussed in this paper.

In this subsection, we show the Li depletion during the post-RGB bump evolution predicted by stellar models. For this purpose, we employed the PAdova and tRieste Stellar Evolutionary Code \citep[PARSEC;][]{2012MNRAS.427..127B}. We computed models of 1\,$M_\odot$ and 1.5\,$M_\odot$ for four metallicities: ${\rm Z=0.0005, 0.001, 0.002, 0.003}$, with $\alpha$-enhanced mixtures \citep[${\rm [\alpha/Fe]=0.4}$;][]{2018MNRAS.476..496F}, corresponding to ${\rm [Fe/H] = -1.7, -1.4, -1.1}$ and $-0.8$\,dex. We also adopted the envelope overshoot efficiency, $\Lambda_{\rm e} = 0.5$ for the 1\,$M_\odot$, while $\Lambda_{\rm e}\!=\!0.7$ for the 1.5\,$M_\odot$; and the same scheme for the core overshoot with $\lambda_{\rm ov}\!=\!0$ and $0.4$ \citep[see][for more information on the standard PARSEC model prescription]{2025A&A...701A.258N}. 

More importantly, we computed these models with thermohaline mixing, which is treated as a diffusive process in PARSEC \citep{2025A&A...696A.136N}. In particular, whence the temperature at the bottom of the convective envelope is sufficient to trigger the ${\rm ^3He(^3He,2p)^4He}$ nuclear reaction, it leads to an inversion of the molecular weight profile and hence establishes the thermohaline instability region \citep[see][and references therein]{2007A&A...467L..15C}. As a star evolves along the RGB phase, the H-burning shell expands outward and eventually overlaps with the thermohaline region, generating a mixing environment that significantly affects the variation of Li. The implementation of thermohaline mixing involves a free coefficient that is the efficiency parameter--$\alpha_{\rm th}$. In this work, we adopted $\alpha_{\rm th}\!=\!50$, calibrated by \citet{2025A&A...696A.136N} using the PARSEC code, for all models. 

\begin{figure}
  \includegraphics[width=1\columnwidth]{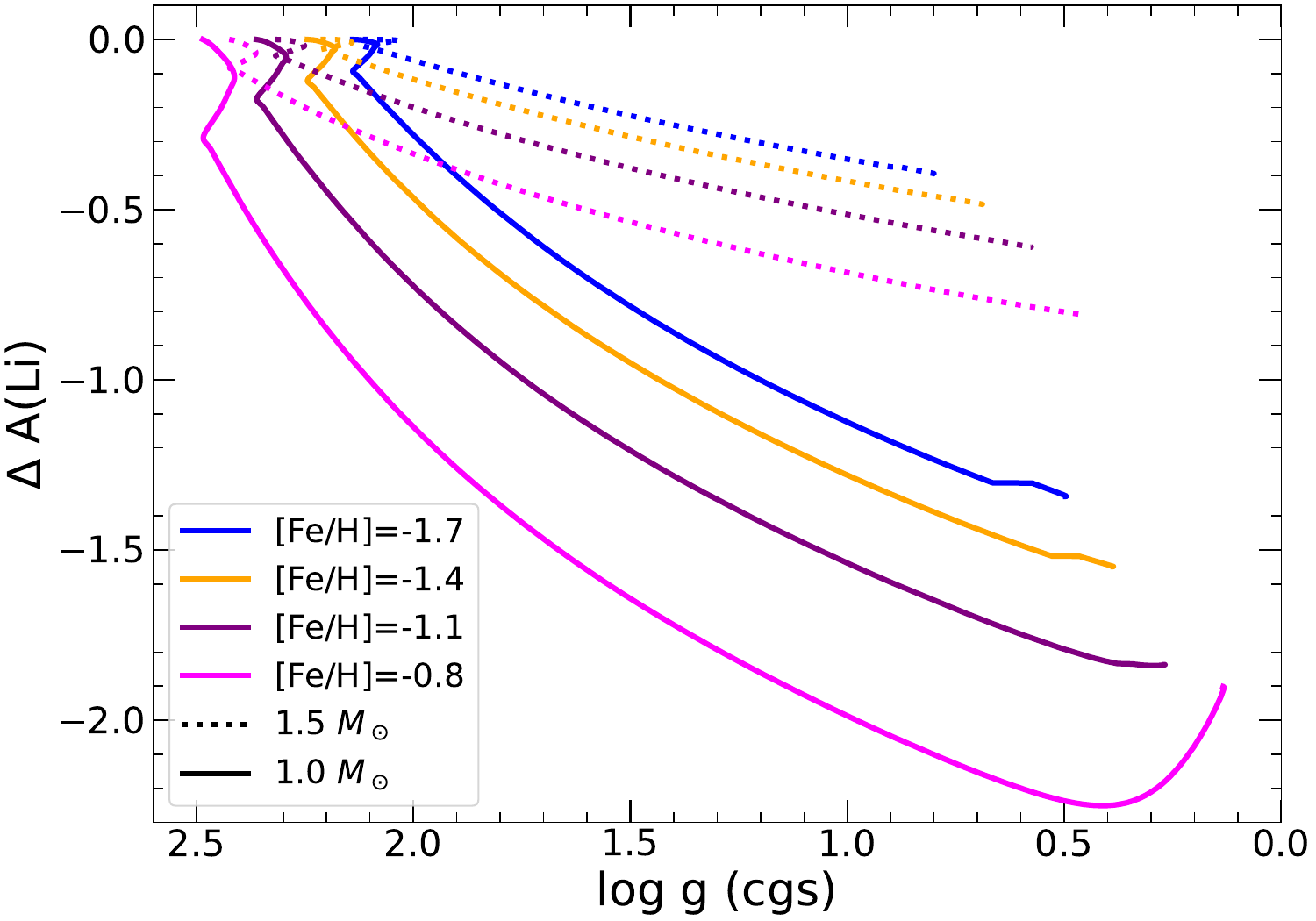}
  \caption{Evolution of ${\rm \Delta A(Li)}$ with $\log g$ during the post-RGB bump evolution of all eight models computed in this work. The definition of ${\rm \Delta A(Li)}$ can be found in the main text.}
  \label{DAli_model}
\end{figure}

While the Spite plateau is observed among the unevolved stars \citep{spite1982}, and the effect of first dredge-up leads to the second Li plateau during the pre-RGB bump phase \citep{mucciarelli2022},  \ali\ during the post-RGB bump phase is greatly varied due to the impact of thermohaline mixing. In Fig.~\ref{DAli_model}, we show the variation of \ali\ with $\log g$ along the post-RGB bump evolution. For clarity, we define ${\rm \Delta A(Li)}$ as the relative \ali\ with respect to its value at the second Li plateau before the RGB bump. First, assuming that all models have the same thermohaline efficiency parameter and the same value of \ali\ at the second plateau, a star with higher metallicity depletes more Li than its lower metallicity counterpart, at the same value of $\log g$. On the other hand, a star with higher mass shows less depletion. This is due to the internal structure differences of stars with different metallicity and mass, which directly relate to the extension of their envelopes. Namely, stars of the same mass and higher metallicity have more extended envelopes, as is also the case for stars of lower mass at the same metallicity. 
This leads to a hint that the scatter in \ali\ during the post-RGB bump evolution observed in the SMC in this work could be due to, at least partially, the stellar mixing process in each individual stars of different metallicity and mass.

Finally, we would like to stress that a Li chemical evolution model of the SMC, taking into account the correction from stellar models would improve the interpretation of our measured \ali\ . This approach has been implemented to study the Li evolution in main sequence and pre-RGB bump stars in the MW thin disc and the Gaia-Sausage-Enceladus dwarf galaxy \citep{nguyen25,2026A&A...710A.224N}. However, such model including stellar correction during the post-RGB bump phase would require a finer grid of stellar models, both in terms of mass and metallicity. 
In other words, the sample presented in this work may serve as a benchmark for calibrating future developing models.

\subsection{Contribution of novae}
\label{ss:novae}

The rise of A(Li) we observe in the MW stars with [Fe/H]$>$-0.8 dex 
is usually ascribable to the contribution of novae, enriching the Galactic interstellar medium on long timescales 
\citep{starrenfield71, dantona1991, romano99, Izzo15, borisov2024, Charbonnel2026}. 
Indeed, chemical evolution models for the Galaxy, predicting the amount of Li formed at any given time, are able to reproduce 
the upper envelope of A(Li) distribution in Galactic dwarf stars by including contribution of novae outbursts
\citep[see e.g.][]{Cescutti19,romano21}.
In the SMC the amount of Li seems to increase with the time, likely due to the contribution of novae, as suggested 
by the Li content of the three GCs of the sample. These clusters in fact allow us to 
have precise ages, at variance with the SMC field stars, and therefore to trace 
the temporal evolution of A(Li) over the entire life of the SMC, as shown in Fig.~\ref{amr} and in Section 4.2.

However, the three SMC GCs sparsely sample the metallicity range as probed by observations (see Fig. \ref{fig_lidstr2}). 
Therefore, to further test the effect of Li enrichment by novae outbursts in terms of metallicity, we run galactic chemical evolution calculations for models of the SMC and the MW disk in the solar neighbourhood.
These models adopt star formation history parameters (e.g. infall timescale $\tau_{inf}$, star formation efficiency $\nu$/star formation timescale $\tau_*$) typically used in the description of the two galactic environments \citep[e.g.][]{Grisoni18,Palla21,Palla26,Escala26}. For what concern the prescriptions for the nova occurrence rates and Li nucleosynthesis, we adopt analogue prescriptions to those assumed in \citet{romano21}. In this way, we anchor to a well-tested nucleosynthetic framework that describe the evolution of Li in the Galactic disk.
A more detailed description of the model equations, features and nucleosynthesis prescriptions is provided in Appendix \ref{app:chem_evo_model}.

\begin{figure}
  \includegraphics[width=0.48\textwidth]{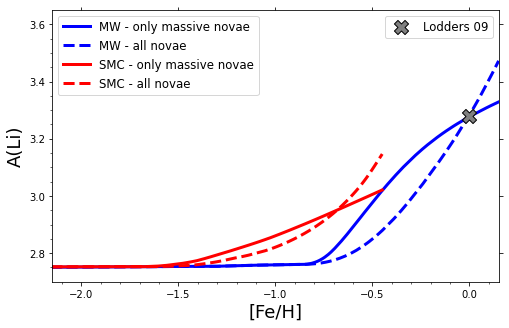}
  \caption{Chemical evolution model for A(Li) as a function of [Fe/H] for MW (blue curves) and 
  the SMC (red curves). In both cases, solid curves are the models including only massive novae and dashed curves 
  the models including all the novae \citep[see e.g.][]{romano21}. 
  Grey cross indicates the meteoritic value of A(Li) \citep{Lodders09}.}
  \label{chemodel}
\end{figure}

In Fig. \ref{chemodel}, we show the predicted A(Li) evolution as function of the metallicity for the models of the MW disk and the SMC, both for the cases where novae contributing to the Li enrichment are in the mass range 1-8 M$_\odot$ (\citealt{romano21}; fiducial model, dashed lines) and 3-8 M$_\odot$ (\citealt{romano21}; alternative model, solid lines). 
It is clear from the figure that the rise in A(Li) starts at different metallicities in the SMC and in the MW disk. This happens irrespectively of the Li enrichment scenario adopted, with the SMC model showing an increase in Li content at [Fe/H]$\gtrsim -1.5 $ dex, while the MW at [Fe/H]$\gtrsim -0.8 $ dex. 

This happens as a natural consequence to the so-called time-delay-model effect (e.g. \citealt{Matteucci12}): the slower pace of star formation of a dwarf galaxy (as the SMC) leads to a slower metal-enrichment rate, resulting in an increased contribution by delayed sources of enrichment (as novae) at lower metallicities relative to what observed in the Galaxy.
Such a theoretical outcome well-align with the abundance data displayed in this paper. Indeed, despite we do not have the intent to reproduce the A(Li) trends displayed in Fig. \ref{fig_lidstr} and \ref{fig_lidstr2} (as models do not take into account Li depletion throughout stellar evolution), the increase in A(Li) from SMC predictions (see red lines in Fig. \ref{chemodel}) falls right in the same metallicity window  as shown by the SMC data ($\rm -1.4\lesssim [Fe/H]/dex\lesssim-0.6$), similarly increasing in A(Li) with metallicity. Moreover, it is worth noting that the bulk of GALAH true measurements (blue dots in Fig. \ref{fig_lidstr}) starts to appear at [Fe/H]$\gtrsim$-0.75 dex. Such a metallicity is very similar to the upturn in A(Li) as predicted by the two models for the MW disc (see blue lines in Fig. \ref{chemodel}). 
This provide further support to the hypothesis that at least part of the observed trend in RGB stars is caused by an increase in the interstellar medium Li content.

\subsection{Rotational velocity}
An additional effect that can contribute to increase the A(Li) star-to-star scatter 
is the stellar rotational velocity.  Several works highlight that A(Li) in giant stars increases 
by increasing the projected rotational velocity \citep[see e.g.][]{carlberg2012, demeiros2000,magrini2021, tsantaki2023}. 
This is a small effect, where the highest rotational velocities are of about 10 km/s. The resolution of the FLAMES-GIRAFFE spectra 
is not sufficient to well measured such slow rotations. 
Moreover, we checked for differences in the broadening of the spectral lines of the observed stars, ascribable to possible high projected rotational velocities. 
The FWHM of each spectrum was measured with the codes DAOSPEC \citep{daospec} and 4DAO \citep{4dao}. 
All the stars, regardless of their \ali\ value, have comparable FWHM, of about 6 km/s, and fully compatible with the typical values for macroturbulence in low-mass stars close to the RGB tip \citep[see e.g.][]{carney2008}, supporting that no significant rotational broadening should be 
invoked to explain the observed line broadening.
However, we cannot exclude that (a marginal) part of the observed \ali\ spread can arise from the rotational velocity distribution of these stars.

\section{Discovery of four Li-rich stars}
\label{lirich}

\begin{figure*}
  \includegraphics[width=1.\textwidth]{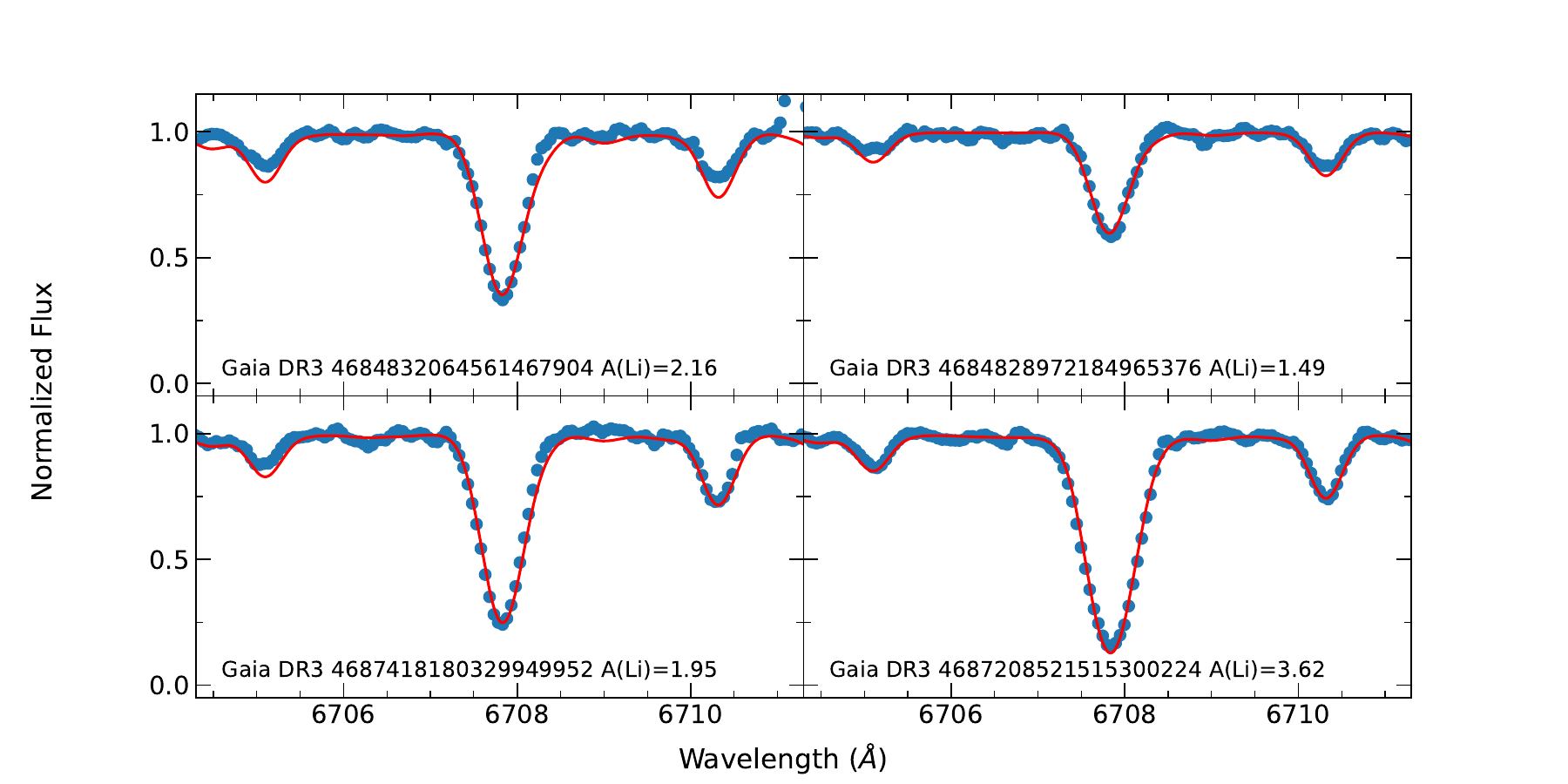}
  \caption{Spectral regions around the Li line at 6707.8 \AA\ for the four Li-rich stars (blue points) 
  with superimposed the best-fit synthetic spectra (red lines).}
  \label{lir_spec}
\end{figure*}

Four stars in our sample have \ali\  greater than 1 dex, significantly higher than those of the rest of the sample. 
Their A(Li) range from +1.49 to +3.62 dex. 
These four stars correspond to $\sim$2\% of the entire sample, consistent with the fraction 
of Li-rich stars detected in our Galaxy \citep{deepak2019,martell2021}.
These are the first low-mass Li-rich stars discovered in the Magellanic Clouds and add to the limited 
list of Li-rich stars identified in external galaxies \citep{kirby2012,hill2019}. 
Figure~\ref{lir_spec} shows the Li line for these four targets, while  
Table~\ref{info2} lists the main information of these stars.

Li-rich stars may form through several different channels, involving either in situ Li production 
or an external origin for the excess Li that is observed. 
The only known internal process capable of producing Li is the Cameron–Fowler mechanism \citep{cameron55,cameron71}. 
It follows an efficient and fast transport process that carries $\rm ^{7}Be$ to cooler layers before it decays into $\rm ^{7}Li$. 
Many Li-rich stars in the MW are found in the red clump \citep{deepak2019,martell2021}, possibly due to the helium flash at the end 
of the RGB that may trigger the Cameron–Fowler mechanism \citep{schwab2020,mori2021}.
An alternative scenario is that Li—again produced through the Cameron–Fowler mechanism—is synthesized in a companion star 
and subsequently transferred through mass transfer onto the observed star. 
AGB stars with initial masses higher than $\sim$3.5--4 $\rm M_{\odot}$ are expected to produce Li 
during the hot bottom burning phase \citep{ventura2011,doherty2014}.

Such a distinction is far from our understanding, since the proposed mechanisms are always neither unambiguous nor available.
If the Li enrichment originates from binary interaction, one may expect the RV variability (although very wide orbits may produce 
variations that are too small or too long-term to be detected),
the enhanced surface rotation as a consequence of angular momentum transfer, 
overabundances of Na \citep{karakas2010, doherty2014} and/or neutron-capture elements produced by the $s$-process \citep{busso1999,cristallo2015}. 
On the other hand, the absence of these signatures does not automatically rule out a binary origin, 
since interaction products may no longer retain all the observable evidence of their binary history.

\begin{figure}
  \includegraphics[width=0.5\textwidth]{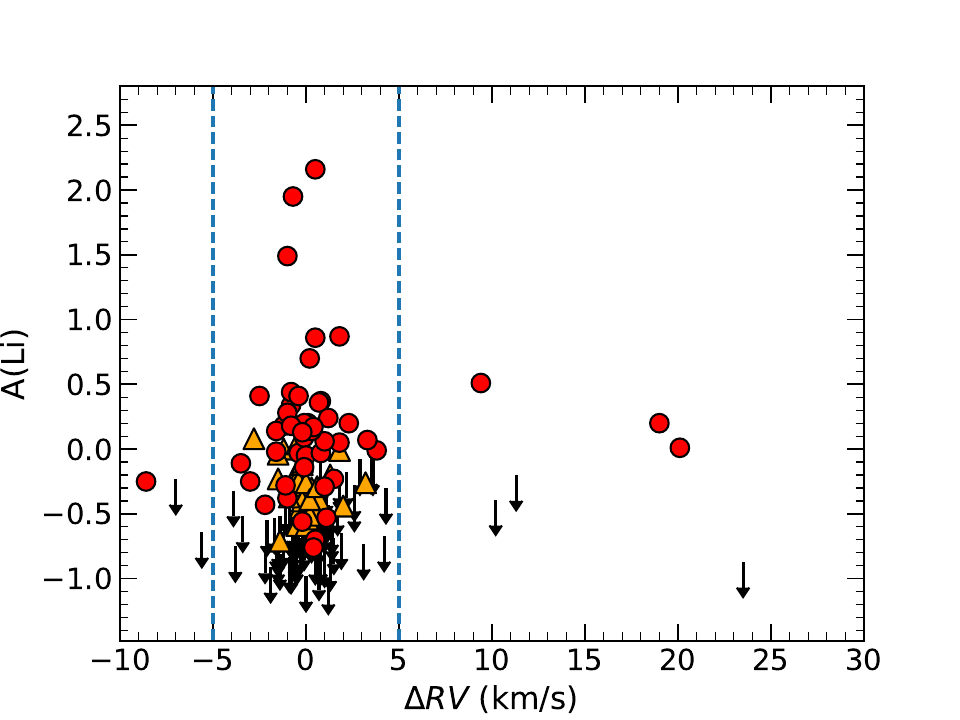}
  \caption{Behaviour of \ali\ as a function of the RV difference measured by \citet{muc23a} and 
  \citet{anoardo} for 158 SMC stars in common with our sample. Symbols are the same as in Fig.~\ref{fig_lidstr}. 
  Stars outside 
  the two vertical dashed lines are binary candidates \citep[see][]{anoardo}.}
  \label{fig_lirv}
\end{figure}

Up to date, the current available spectroscopic datasets do not allow us a comprehensive investigation 
of all the proposed mechanisms that can discriminate between an internal and an external origin of Li enrichment. 
For 158 targets of our sample, RVs are available from two epochs \citep{muc23a,anoardo}.
\citet{anoardo} identified nine stars with a RV variation between the two epochs compatible 
with binary systems ($>$5 km/s), while for the other stars the RV variations are compatible 
with the RV variations arising from radial pulsation. 
Fig.~\ref{fig_lirv} shows \ali\ as a function of the RV difference 
between the two epochs. No particular relationship is observed between \ali\ and RV variation. 
The stars considered as binary candidates do not show peculiar \ali\ values, nor do they preferentially belong 
to either the upper-limit sample or the sample with actual measurements. 
Moreover, for three of the stars with \ali $>$ 1 dex there are two RV epochs available, but these stars 
are not binary candidates, suggesting that these Li-rich objects are not associated with binary systems. 
Furthermore, three Li-rich stars have RUWE values around 1.0, indicating a well-behaved single-star astrometric solution, 
while the star Gaia DR3 4687418180329949952 has RUWE=1.70, suggesting a possible 
binary nature, despite the lack of a significant RV variation.

The abundances of Na and $s$-process elements can help constrain the initial mass of a putative AGB companion. 
Low-mass AGB stars ($\sim$1–3 $\rm M_{\odot}$) 
can efficiently produce second-peak $s$-process elements, such as Ba and La, particularly at low metallicity. 
In intermediate-mass AGB stars, the production of heavy $s$-process elements is generally less efficient, 
while first-peak elements, such as Y and Zr, tend to become relatively more prominent.
In super-AGB stars, $s$-process enrichment is generally limited, whereas substantial Na production may occur through hot-bottom burning. 
The Li-rich stars with [Fe/H]$<-$3.0 dex are characterized by a strong enhancement of Na \citep[see Fig. 6 in][]{mucciarelli24}. 

For the stars in common with \citet{muc23a}, we can compare \ali\ with the abundances of Na, Zr, Ba and La, 
because these elements can provide constraints on the mass of the AGB companion in the binary scenario.
We find no correlation between \ali\ and the abundances of Na and $s$-process elements. 
In particular, \citet{muc23a} identified a number of stars exhibiting significant enhancements in Ba and La, 
suggesting that they may be products of mass transfer in binary systems, but no correlation is detected 
between A(Li) and these abundances.
Specifically, the three Li-rich stars in common with \citet{muc23a} show Na, Zr, Ba and La abundances that are fully consistent 
with those of the rest of the sample, whereas stars exhibiting significant Zr, Ba and La enhancements 
display normal \ali\ .

\begin{table*}[h!]
\caption{Information about the SMC Li-rich stars.}
\label{info2}
\centering
\begin{threeparttable}
\begin{tabular}{cccccccccc}
\toprule
ID Gaia DR3 & [Fe/H] & \teff\ &  \grav\  &\ali &  $\rm EW_{Li}$ & RUWE & $\delta_{\rm RV}$ & Field & $\rm ID_{\rm Mucciarelli+23a}$ \\
 & (dex) & (dex) &  (K) & (dex) & (m\AA) & & (km/s) & & \\
\midrule
4684832064561467904  &  $-$0.52 & 4405 &  0.93 &      2.16 & 373.6 & 1.06  &  +2.5  & FLD-339 &100052 \\
4684828972184965376  &  $-$1.10 & 4378 &  0.90 &      1.49 & 211.8 &  1.00  &  $-$1.0  & FLD-339 & 101785  \\
4687418180329949952  &  $-$0.97 & 3983 &  0.50 &      1.95 & 457.9 &  1.70  &  $-$1.3  & FLD-419 &100857 \\
4687208521515300224  &  $-$1.14 & 4018 &  0.55 &     3.62 & 565.5 &  1.00  &  --  &   FLD-419 & --- \\
\bottomrule
\end{tabular}
\begin{tablenotes}
\small
\item \textbf{Notes.} Columns are: ID from Gaia DR3 \citep{vallenari23}, 
[Fe/H] from \citet{anoardo}, \ali\ and EW of Li line, RUWE parameter from Gaia DR3, difference of RV from \citet{anoardo} and \citet{muc23a}, 
field of observation and ID from \citet{muc23a}.
\end{tablenotes}
\end{threeparttable}
\end{table*}

As discussed above, we checked possible significant differences in the broadening of the observed lines, as hints of 
high rotational velocity. The four Li-rich stars have FWHM fully compatible with those of the other targets.
In summary, one of the four Li-rich stars could be explained by the binary system scenario 
which is supported by its high RUWE value; whereas, both scenarios remain possible for the other three stars.


\section{Discussion and conclusions}
\label{end}

This work presents the first determination of \ali\ in low-mass, post-RGB bump stars 
in the SMC, among  other information about A(Li) in this galaxy arising, e.g., from interstellar medium 
\citep{howk12,molaro24} and massive AGB stars \citep{Hatzidimitriou2003}.
The main results of this work are summarized as follows:
\begin{itemize}
\item 
The Li line is clearly detected at a statistical significance higher than 5$\sigma$ in 32\% of the sample, 
a fraction that increases to 53\% when stars with detections between 3$\sigma$ and 5$\sigma$ are also included. 
The main feature of A(Li) distribution in SMC stars of our sample is a significant star-to-star scatter 
for stars with [Fe/H]$>$--1.4 dex that cannot be explained by the observational uncertainties. 
Additionally, the bulk of our stars dominates at a metallicity range that is significantly lower than those of the MW stars.
\item
Concerning the significant star-to-star scatter observed in our sample, it is thought mainly due to the extra mixing during this post-RGB bump phase, 
which is expected to depend on stellar mass and metallicity. The SMC hosts stars spanning 
a wide range of ages and, consequently, stellar masses. 
Models including thermohaline mixing as the mechanism responsible for the extra mixing provide 
an illustrative example, showing that variations in stellar mass and metallicity can account for 
a large fraction of the observed A(Li) scatter. 
\item
In the latter finding, we anticipate that the SMC has experienced a lower star formation efficiency than the MW \citep{hz04,Rubele2018,Palla26}, 
therefore the increase in A(Li) 
driven by the contribution of novae is expected to occur at lower metallicity than in our Galaxy. 
SMC-tailored chemical evolution models predict this rise should start at approximately [Fe/H]$\sim$--1.5 dex, namely already at lower metallicities than those measured in our RGB sample. 
\item 
The presence of these two major mechanisms (the thermohaline mixing and the contribution of the novae) 
complicates greatly the interpretation of the variation of A(Li).
This prevents us to infer the initial A(Li) of the these stars and investigate whether the SMC shares the same initial 
Li content as the MW.
\item
The A(Li) measurements obtained for stars in three SMC globular clusters provide temporal constraints 
on the evolution of Li in the galaxy. For the two oldest clusters, with ages between 6 and 10 Gyr, 
we derive only upper limits, A(Li)$<-$0.3/--0.5 dex. In contrast, 
the younger cluster NGC 419 (1.5 Gyr) exhibits a higher \ali\ , reaching A(Li)$\sim$+0.3 dex. 
This result suggests that \ali\ in the SMC could increase over time, most likely owing 
to the relatively recent contribution of novae. However, at young ages, a large A(Li) 
dispersion could be present (similar to that observed in the MW disk stars)
Measuring \ali\ in other 1-2 Gyr old SMC GCs will therefore be crucial for constraining the degree 
of chemical homogeneity produced by nova enrichment.
\item 
Finally, we identified four stars with \ali\ $\geq$1.0 dex, significantly higher than the other stars in the 
sample and that we labelled as Li-rich stars. These are the first low-mass Li-rich stars discovered so far 
in the SMC. The data in hand do not allow to disentangle between an internal scenario where the fresh Li 
is produced in-situ and an external scenario where the fresh Li is produced in a massive companion 
in a binary system. 
Only one of these Li-rich stars has a RUWE parameter that suggests 
the external scenario. However, no RV variation is detected between the two available epochs. 
\end{itemize}

All the findings we obtained from this spectroscopic dataset demonstrate 
that also the \ali\ we measured in RGB stars close to the RGB tip 
still provide important and valuable information and constraints on the chemical 
evolution of Li in galaxies with star formation histories different than 
that of the MW.

\begin{acknowledgements}
The authors thank the referee, Grazina Tautvaisiene, for the constructive comments. 
A.M. thanks Piercarlo Bonifacio for the useful discussions.
Co-funded by the European Union (ERC-2022-AdG, "StarDance: the non-canonical evolution of stars in clusters", Grant Agreement 101093572, PI: E. Pancino).
Views and opinions expressed are, however, those of the author(s) only and do not necessarily reflect those of the European Union 
or the European Research Council. Neither the European Union nor the granting authority can be held responsible for them.  
L.M. gratefully acknowledges support from ANIDFONDECYT Regular Project n.1251809. 
\end{acknowledgements}

\bibliographystyle{aa} 
\bibliography{biblio} 

\appendix

\section{A(Li) in GALAH DR4 post-RGB bump stars}
\label{app:galah}

\begin{figure}[h]
  \includegraphics[width=0.5\textwidth]{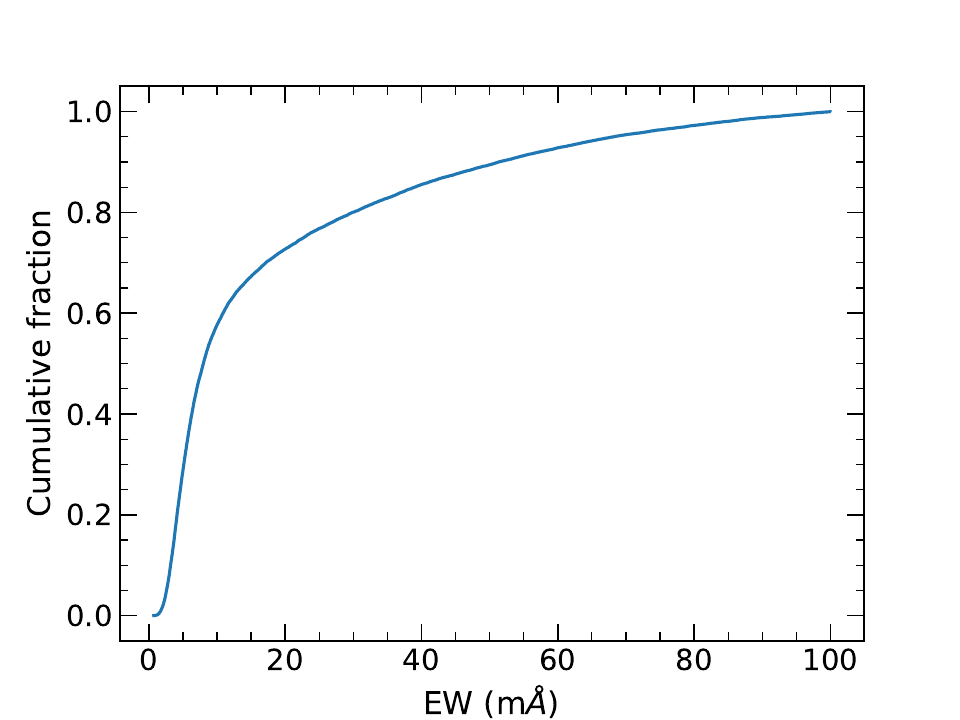}
  \caption{Cumulative distribution of the EW of Li line in the post-RGB bump stars 
  in GALAH DR4 \citep{wang2024}.}
  \label{app_galah1}
\end{figure}

We check the distribution of the EW of Li line as reported in the value-added catalogue
galah\_dr4\_vac\_3dnlte\_a\_li\_240705.fits \citep{wang2024}. 
We restrict our check to post-RGB bump stars with \teff\ $<$ 5000 K, log~g$<$2.0, and 
for which A(Li) values derived from actual measurements are reported 
Figure~\ref{app_galah1} shows the cumulative distribution of the EWs of Li line for these stars. 
Approximately 60\% of the stars have EW$<$10 m\AA , while 70\% have EW$<$20 m\AA . Such small EWs can be reliably measured
only in spectra with adequate S/N ratio and in isolated lines, or at least in lines not significantly affected by blending. 
This is particularly relevant for the Li line in high-metallicity post-RGB bump stars, with solar or super-solar metallicities, 
as is the case for Li line in most stars in GALAH DR4.

\begin{figure}[h]
  \includegraphics[width=0.5\textwidth]{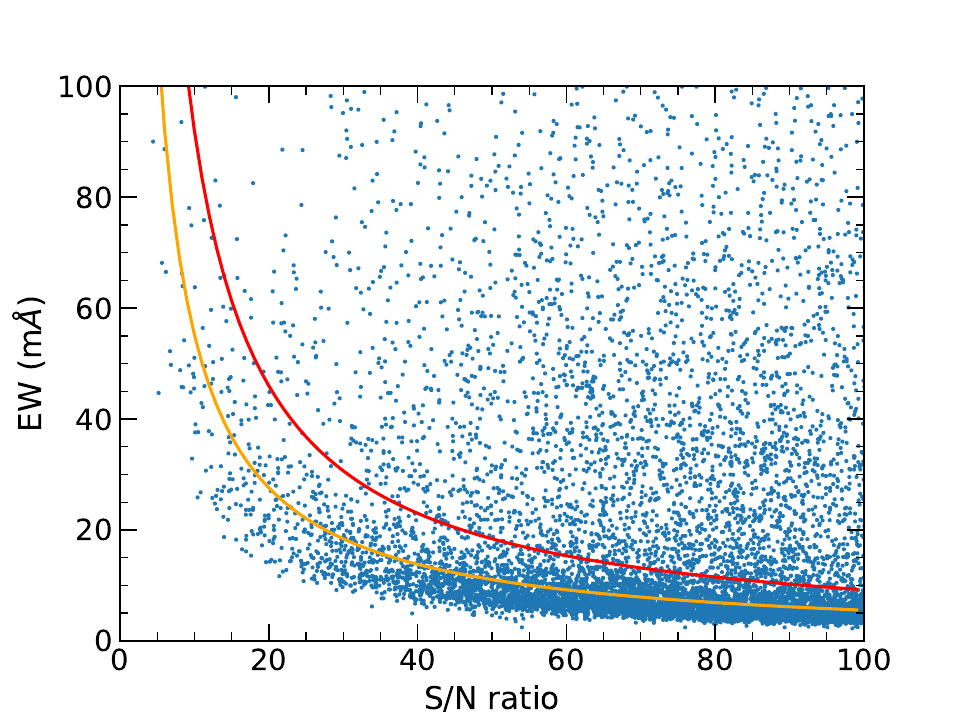}
  \caption{Behaviour of the Li line EW as a function of S/N ratio for the same stars
  shown in Fig.~\ref{app_galah1}. Orange and red lines indicate 3 and 5$\sigma_{Cayrel}$, 
  respectively.}
  \label{app_galah2}
\end{figure}

Figure~\ref{app_galah2} shows, for reference, the distribution of EW as a function of the S/N ratio of the corresponding spectrum. 
The red and orange curves indicate three and five times the uncertainty associated with the EW, estimated using the formula by \citet{cayrel88}, 
as a function of S/N ratio. 
We assumed 5$\sigma_{Cayrel}$ as a robust threshold for the line detectability. 
It is important to keep in mind, however, that the formula by \citet{cayrel88} 
assumes an isolated line profile, an assumption that does not hold for many GALAH stars.

\section{Chemical evolution of Lithium}
\label{app:chem_evo_model}

In Section \ref{ss:novae} we model the chemical evolution of Li in the SMC and the solar neighbourhood to give a first assessment of the contribution of novae to A(Li) trends observed. Here we present in more detail the models adopted.  
In Section \ref{apps:chem_evo} we briefly summarize the main features (equations, parameter assumptions) of the chemical evolution models adopted for the two systems, whereas in Section \ref{apps:nucleos} we focus on the nucleosynthetic prescriptions adopted in such models, with a particular focus on Li.

\subsection{Chemical evolution models}
\label{apps:chem_evo}

The chemical evolution models discussed in this work rest on the following basic equation that describes the chemical evolution of a given element $i$ (see, e.g. \citealt{Matteucci12}):
\begin{equation}
    \Dot{G}_i (t) = -\psi(t)\, X_i(t)\, + \,R_i(t)\, +\, \Dot{G}_{i,inf}(t)\, -\, \Dot{G}_{i,out},
    \label{eq:basic_chemevo}
\end{equation}
where $G_i(t) = X_i(t)\,G(t)$ is the fraction of the gas mass in the form of an element $i$ and $X_i(t)$ represents the abundance fraction in mass of a given element $i$.

The first term on the right-hand side of Eq. \eqref{eq:basic_chemevo} corresponds to the rate at which an element $i$ is removed from the interstellar medium due to star formation.  The star formation rate ($\psi(t)$, hereafter SFR) is parametrised according to the Schmidt-Kennicutt law (\citealt{Kennicutt98}), with the star formation efficiency (SFE) $\nu$ as the control parameter that represents the SFR per unit mass of gas. 
For the solar neighbourhood model (hereafter MW model) and the SMC model, two different values of $\nu$ are adopted: in particular, we set $\nu=1 \, \rm Gyr^{-1}$ for the MW and $\nu=0.1 \, \rm Gyr^{-1}$ for the SMC, in agreement with assumptions in previous chemical evolution studies (e.g. \citealt{Grisoni18,Palla21,romano21} for the the solar vicinity; e.g. \citealt{Palla26,Escala26} for the SMC). 

$R_i(t)$ takes into account the nucleosynthesis from different stellar sources, weighted according to the initial mass function (IMF, here from \citealt{Kroupa93}). This term also includes the products originating from binary systems such as as Type Ia SNe.
For the latter, we assume the delay-time-distribution (DTD) by \citet{MatteucciRecchi01}, which enables us to obtain abundance patterns that are similar to those obtained with other literature Type Ia SNe DTDs (see, e.g. \citealt{Palla21} for details).

The last two terms of Eq. \eqref{eq:basic_chemevo} refer to the gas flows history of the galaxy, namely inflows and outflows.
We model the rate of gas inflow in terms of an exponentially-decaying law (see, e.g. \citealt{Chiappini97,Chiappini01}):
\begin{equation}
    \Dot{G}_{i,inf} \propto X_{i,inf}\, e^{-t/\tau_{inf}}, 
    \label{eq:infall}
\end{equation}
where $X_{i,inf}$ is the chemical abundance of the element $i$ of the infalling gas, here assumed to be primordial, and $\tau_{inf}$ the infall timescale. As for the SFE, $\tau_{inf}$ is set to depend on the specific history of star formation of the galaxy: for MW model we assume $\tau_{inf}= 7 \, \rm Gyr$ (see e.g. \citealt{Grisoni18,romano21} and references therein), whereas for the SMC $\tau_{inf}= 10 \, \rm Gyr$ (as already assumed to model fairly massive dwarf galaxies, see e.g. \citealt{Gioannini17,Palla20a}).

For what concerns galactic winds, we assume them to be proportional to the SFR (e.g. \citealt{Vincenzo15,Palla20b,Escala26}):
\begin{equation}
    \Dot{G}_{i,out} = \omega_i \, \psi(t),
    \label{eq:outflow}
\end{equation}
where $\omega_i$ is the mass loading factor parameter, assumed to be the same for all chemical elements $i$. Here, we assume $\omega_i = 0$ for the MW (see \citealt{Meioli09,Spitoni09,Hopkins23}) and  $\omega_i = 0.25$ for the SMC (see \citealt{Gioannini17,Palla20a}). 

It is worth noting that the parameter sets adopted in this work allow us to well reproduce the measured rates of SF in both the MW disk and the SMC. In particular, for the MW we obtain a present-day rate of 1.95 M$_\odot$ yr$^{-1}$ ($1.9\pm0.4$ M$_\odot$ yr$^{-1}$ in \citealt{Chomiuk11}), while for the SMC of 0.059 M$_\odot$ yr$^{-1}$ ($0.06^{+0.02}_{-0.01}$ M$_\odot$ yr$^{-1}$ average over the last 100 Myr in \citealt{Rubele2018}).

\subsection{Nucleosynthesis prescriptions}
\label{apps:nucleos}

All the galaxy models adopted in this work adopt the same sets of nucleosynthesis prescriptions.
Regarding the stellar yields from "standard" sources, we adopt those by 
\citet{Ventura13,Ventura20} for low- and intermediate-mass stars 
, \citet{Nomoto13} (\citealt{WW95} in the case of Li\footnote{\citet{WW95} are the only yield set to date offering a grid that span a range of stellar masses and metallicities adequate for chemical evolution that account for the $\nu$-process effect for the nucleosynthesis of Li (see \citealt{romano21} for more details on this implementation).}) for massive stars, and \citet{Iwa99} for Type Ia SNe. 

For what concerns Li, we tested either the Spite plateau value, $\rm A(Li)_{obs,P}\sim2.2$ dex, or
the BBN-predicted value, $\rm A(Li)_{obs,P}\sim2.7$ dex (from Planck, see \citealt{Pitrou18} and references therein), 
as the initial \ali\ in the chemical evolution models. However, in the main text we show results only for models with the BBN initial abundance, as analogue trends with metallicity are obtained with the two setups and we are not interested in the predicted absolute abundances.
For the sources of production of Li, in addition to "standard" stellar sources (contributing negligibly to Li enrichment, see also \citealt{Cescutti19}), the model consider Li nucleosynthesis from classical novae and Galactic cosmic-ray (GCR) spallation process, following the prescriptions described in \citet[][and references therein]{romano21}. 
For classical novae, we tested two different scenarios of enrichment, taken from \citet{romano21}. These are chosen as they are calibrated to obtain the current nova outburst rate in the Galaxy \citep{Izzo15} as well as meteoritic A(Li) \citep{Lodders09} at solar metallicity (see also Figure \ref{chemodel}), still being compatible theoretical and observational probes on the Li production by these objects \citep{Izzo15,Starrfield20}: 
\begin{enumerate}
    \item {\it fiducial model}: we assume that $\sim$2\% of all stars with initial mass between 1-8 M$_\odot$ enter the formation of nova systems. For such systems, the average Li production per nova outburst is set to $2.55 \times 10^{-10} \, \rm M_\odot$. Accounting for the repeated outburst of novae systems (10$^4$, e.g. \citealt{Bath98}), this results in a total ejected mass of Li per nova of 2.5$\times 10^{-6} \, \rm M_\odot$;
    \item {\it alternative model}: we assume $\sim$20\% of all stars with initial mass between 3-8 M$_\odot$ are summed to form novae systems. In this case, the total ejected mass of Li per nova (accounting for the repeated outbursts) is of 1.45$\times 10^{-6} \, \rm M_\odot$. 
\end{enumerate}
It is worth noting that for both the scenarios tested we obtain a SMC nova outburst rate of $0.5$ event/yr, in line with the one measured in the SMC ($0.9\pm0.4$ event/yr, \citealt{Mroz16}). 

Finally, cosmic-ray bombardment of the interstellar medium is also a (secondary) factory of Li in galaxies. Here, the contribution of GCR spallation is taken into account by incorporating the absolute yields in \citet[][see their Table 1, lower-bound spectrum]{Lemoine98}, as in \citet{romano21}.

\end{document}